\documentclass[epj]{webofc}
\usepackage[varg]{txfonts}   
\usepackage{color}
\usepackage{epsfig}
\woctitle{Mathematical Modeling and Computational Physics 2015}

\usepackage[english]{babel}
\begin{document}
\title{Numerical calculation of scaling exponents of percolation
process in the framework of  renormalization group approach}
%
%

\author{L.~Ts.~Adzhemyan\inst{1} \and
        M.~Hnati\v{c}\inst{2,3,4}\fnsep\thanks{\email{hnatic@saske.sk}} \and
        M.~Kompaniets\inst{1} \and
        T.~Lu\v{c}ivjansk\'{y}\inst{2, 5} \and
        L.~Mi\v{z}i\v{s}in\inst{2,4}\fnsep\thanks{\email{lukas.mizisin@student.upjs.sk}}
}

\institute{Department of Theoretical Physics, St. Petersburg University,\\
	Ulyanovskaya 1, 198504 St. Petersburg-Petrodvorets, Russia
\and	
	Department of Theoretical Physics, \\ 
	SAS, Institute of Experimental Physics, \\
	Watsonova 47, 040 01 Ko\v{s}ice,  Slovakia 
\and
           Bogoliubov Laboratory of Theoretical Physics, \\
           Joint Institute for Nuclear Research, \\
           Joliot-Curie 6, 141980 Dubna, Moscow Region, Russia 
\and
           Faculty of Science, P.~J.~\v{S}af\'arik University, \\ 
           \v{S}rob\'arov\'a 2, 041 54 Ko\v{s}ice, Slovakia
\and		
		   Fakult\"{a}t f\"{u}r Physik, Universit\"{a}t Duisburg-Essen,\\
		   D-47048 Duisburg, Germany
          }

\abstract{%
  We use the renormalization group theory to study the directed bond percolation (Gribov process) near its second-order phase 
  transition between absorbing and active state. We present a numerical calculation of  the
  renormalization group functions in the $\epsilon$-expansion where $\epsilon$ is a deviation from the 
  upper critical dimension $d_c = 4$. Within this procedure anomalous dimensions $\gamma$  are expressed in 
  terms of irreducible renormalized Feynman diagrams and thus the calculation of renormalization constants
  could be entirely skipped. The renormalization group is included by means of the $R$ operation, and for 
  computational purposes we choose the null momentum subtraction scheme. 
}
\maketitle
%
{\section{Introduction} \label{intro}}
The renormalization group (RG) method is a theoretical framework which is especially suitable for
studying various critical phenomena \cite{Vasiliev,Amit}. From a computational point of view it provides techniques for a 
perturbative calculation of different critical exponents. One of the most prominent dynamical models \cite{Tauber2014} which
 exhibits a second order phase transition is  the directed bond percolation \cite{Stauffer,HHL08}.
%
%
 In the physical literature it is known also as Schl\"ogl first reaction \cite{Schlogl}. 
 Among others it explains hadron interactions at very high
 energies (Reggeon field theory) \cite{Cardy}, stochastic reaction-diffusion processes on a lattice \cite{Hinrichsen}, spreading 
 of infection diseases \cite{ Janssen}, etc.   
The critical exponents are calculable in the form of  perturbative expansion in a formally small parameter $\epsilon$. 
We note that two-loop results for the exponents $z$ and $\delta$ were obtained in \cite{Bronzan} and 
exponents $\nu$ and $\beta$ later on in \cite{Janssen81}. All necessary information concerning percolation process
in terms of reaction-diffusion model can be found in
the review article \cite{Janssen}.
%
%

%
%
As is in detail discussed in the literature \cite{Vasiliev} (Part 3.5) or
\cite{Amit} (Part 7.5) the central idea behind renormalization group
is freedom in choose of particular renormalization scheme. All of them
makes a theory finite with respect to ultraviolet divergences and regarding universal quantities they
lead to the same result. For practical numerical calculations it is more convenient
 to choose the subtractions at normalization point $\{p=0, \omega=0, \tau=\mu^2\}$
 as explained in \cite{Vasiliev} (see Eq. (3.18)).

%
%

Furthermore, in the renormalized Green function it is possible to replace an additional
contribution of the renormalized constant by the operator $R$ \cite{Vasiliev} applied to Green functions
\begin{equation}
  \Gamma^{R} = R \Gamma = (1-K)R' \Gamma,
  \label{eq-R_operator}
\end{equation}
where $R'$ is the  incomplete operator $R$ that cancels divergences in subgraphs of a
given graph and the operator  $(1-K)$ eliminates the remaining superficial divergence. 

%
%
  The main of this work is to show main steps of alghoritmic procedure, which allows us
  to reproduce known two-loops results to very high precision. Moreover it easy to generalize
  our procedure to high orders and thus obtaine more reliable results.
%
%
{ \section{Renormalization of the model} \label{sec-1}}
A field theoretical formulation of the percolation process \cite{Janssen} is based on
 the following action 
\begin{equation}
  S = \psi^{\dagger}(-\partial_t + D_0  \partial^2-D_0\tau_0)\psi+
  \frac{D_0 \lambda_0}{2}[(\psi^{\dagger})^2 \psi -\psi^    {\dagger} \psi^2],
  \label{eq-actionDP}
\end{equation}
where $\psi$ is a coarse-grain density of percolating agents, $\psi^{\dagger}$ is an auxiliary
 (Martin-Siggia-Rose) response field, $D_0$ is a diffusion constant, $\lambda_0$ is a positive coupling
  constant and $\tau_0$ is a deviation from the threshold value of injected probability (an analog
  of critical temperature in static models). The model is studied near its critical dimension $\epsilon = 4 - d$ in the  region where $\tau_0$ acquires its critical value. The expansion parameter of the perturbation theory is rather $\lambda_0^2$ than $\lambda_0$ as it could be easily seen by a direct inspection of Feynman diagrams. 
  Hence it is more convenient to introduce a new charge $u=\lambda^2$.
The renormalized action functional can be written in the following form
\begin{equation}
  S_R = \psi^{\dagger}(-Z_1\partial_t + Z_2 D  \partial^2- Z_3D\tau)\psi+
  \frac{Z_4D \lambda \mu^{\epsilon}}{2}[(\psi^{\dagger})^2 \psi -\psi^{\dagger} \psi^2].
  \label{eq-actionDP_renorm}
\end{equation}
It can be shown  \cite{Janssen} that this kind of a model is multiplicatively renormalizable. Furthermore
 the action functional $S_R$ can also be  obtained from the action $S$ by the standard 
  procedure of multiplicative renormalization of all the fields and parameters
\begin{eqnarray}
  \psi_0 = \psi Z_{\psi}, \hspace{0.5cm} 
  \psi_0^{\dagger} = \psi^{\dagger} Z_{\psi^{\dagger}}, \hspace{0.5cm}
  D_0 = D Z_D, \hspace{0.5cm}
  \lambda_0 = \lambda \mu^{\epsilon} Z_{\lambda}, \hspace{0.5cm}
  \tau_0 = \tau Z_{\tau}.
\end{eqnarray}
The relations between renormalized constants $Z_i, i=1,2,3,4$ are obtained
in a straightforward fashion and read
\begin{align}
  Z_1 &= Z_{\psi} Z_{\psi^{\dagger}},  
  &Z_2& = Z_D Z_{\psi} Z_{\psi^{\dagger}},\nonumber \\
  Z_3 &= Z_D Z_{\tau} Z_{\psi} Z_{\psi^{\dagger}}, 
  &Z_4& = Z_D Z_{\lambda} Z_{\psi^{\dagger}}^2 Z_{\psi} = Z_D Z_{\lambda} Z_{\psi^{\dagger}} Z_{\psi}^2.
  \label{eq-ren_const}
\end{align}
Moreover, the relation $Z_{\psi} = Z_{\psi^{\dagger}}$ is  satisfied.
  In this work, at the normalization point (NP) $p=0$, $\omega=0$ and $\tau=\mu^2$ is
considered. The counterterms are then specified at the normalization point (NP),  and it is advantageous
 to express renormalization constants in terms of normalized Green functions
\begin{align}
  &\bar{\Gamma}_1 = \partial_{i\omega} \Gamma_{\psi^{\dagger}\psi}\big|_{p=0,\omega=0}, 
  &\bar{\Gamma}_3& =-\frac{\Gamma_{\psi^{\dagger}\psi}-
  \Gamma_{\psi^{\dagger}\psi}\big|_{\tau=0}}{D \tau} \Big|_{p=0,\omega=0}, \nonumber \\
  &\bar{\Gamma}_2 =-\frac{1}{2D}\partial_{p}^2 
  \Gamma_{\psi^{\dagger}\psi}\big|_{p=0,\omega=0},
  &\bar{\Gamma}_4& =\frac{\Gamma_{\psi^{\dagger}\psi^{\dagger}\psi}-
  \Gamma_{\psi^{\dagger}\psi\psi}}{D \lambda \mu^{\epsilon}}    \Big|_{p=0,\omega=0},
  \label{eq:norm_green}
\end{align}
that satisfy the following conditions
$
  \bar{\Gamma}_i^R|_{\tau=\mu^2} = 1,\quad i=1,2,3,4.
$
%
%
RG constants defined by these conditions 
do not depend on $m$, like in minimal
subtraction (MS) scheme. Accordingly RG equations are the same as in MS
scheme

%
%
\begin{equation}
(\mu \partial_{\mu}+\beta_u\partial_u-\tau\gamma_{\tau}\partial_{\tau}-D\gamma_D\partial_D) \Gamma_i^R = (n_{\psi}\gamma_\psi+n_{\psi^{\dagger}}\gamma_{\psi^{\dagger}})\Gamma_i^R,
\end{equation}
where $\mu$ is a reference mass scale, $n_{\psi}$ and $n_{\psi_{\dagger}}$ are the numbers of the corresponding fields
 entering the  Green function under consideration, $\gamma_x = \mu \partial_{\mu} \log Z_x$ are anomalous
dimensions and $\beta_u=u(-2\epsilon-\gamma_u)$ is a beta function describing a flow of the charge $u$ under the
RG transformation \cite{Vasiliev}.
  Using these equations we find relations for the normalized functions
\begin{equation}
  (\mu \partial_{\mu} + \beta_u\partial_u - \tau\gamma_{\tau} \partial_{\tau} - D\gamma_D\partial_D) 
  \bar{\Gamma}_i^R =     \gamma_{i}\bar{\Gamma}_i^R.
\end{equation}
Here, anomalous dimensions $\gamma_i$ are obtained from  relations (\ref{eq-ren_const}) between the renormalization constants 
\begin{equation}
   \gamma_1 = 2 \gamma_{\psi}, \quad \gamma_3  = 2\gamma_{\psi} +\gamma_D +\gamma_{\tau}, \quad
   \gamma_2 = 2 \gamma_{\psi} + \gamma_D,\quad \gamma_4 = 3 \gamma_{\psi} + \gamma_D +  \gamma_{\lambda}.
   \label{eq-gamma}
\end{equation} 

Taking into account the renormalization scheme we can express the anomalous dimension  in terms of the  renormalized 
derivatives of the one-particle irreducible Green function $\bar{\Gamma}_i$ at the normalization
point \cite{Adzhemyan2011, Adzhemyan2013, Adzhemyan2014}
\begin{equation}
F_i \equiv - [\tilde{\tau} \partial_{\tilde{\tau}} \bar{\Gamma}_i^R (\tilde{\tau})]\big|_{\tilde{\tau}=1},
 \quad i =1,2,4
\label{eq-F}
\end{equation}
where $\tilde{\tau} = {\tau}/{\mu^2}$. At the normalization point ($\tilde{\tau}=1$), $\gamma_i$ 
takes the form \cite{Adzhemyan2011,Adzhemyan2013}
\begin{equation}
\gamma_i = \frac{2F_i}{1+F_2-F_3},\quad i =1,2,4.
\label{eq-gamma_F}
\end{equation}
 For  later considerations  it is reasonable to introduce new functions   (see \cite{Adzhemyan2013})
\begin{equation}
f_i \equiv R [-\tilde{\tau} \partial_{\tilde{\tau}} \bar{\Gamma}_i (\tilde{\tau})]\big|_{\tilde{\tau}=1}.
\label{eq-f}
\end{equation} 
These functions are related to the functions $F_i$  (\ref{eq-F})  in the following way  
\begin{equation}
f_i - F_i = f_i F_3, \hspace{1cm} i=1,2,4.
\label{eq-f_F}
\end{equation}
We rewrite equations in (\ref{eq-gamma_F}) to obtain relations for anomalous dimensions  in terms of the  renormalized derivatives 
of the one-irreducible Green function $\bar{\Gamma}_i$ with respect to $\tilde{\tau}$ at the normalized point
\begin{equation}
\gamma_i = \frac{2 f_i}{1+ f_2}, \quad i=1,2,4.
\end{equation}
The main benefit of this procedure considering (\ref{eq-f}) is that the operator $R$ is taken at
the  normalization point and it can be expressed in terms of a subtracting 
operator $1-K_i$  that  eliminates all divergences from 
the Feynman graphs \cite{Vasiliev}
\begin{equation}
R \Gamma = \prod\limits_i (1-K_i)\Gamma,
\label{eq-Roperator}
\end{equation} 
where the product is taken over all relevant subgraphs of the given Feynman graph, including also the graph as a whole. 
In the NM scheme we obtain the  following representation  for the $R$-operator
 \cite{Adzhemyan2011, Adzhemyan2014}
\begin{equation}
R \chi =\prod\limits_{i} \frac{1}{n_i!}\int\limits_0^1 da_i (1-a_i)^{n_i} \partial^{n_i+1}_{a_i} \chi(\{a\}),
\end{equation}
where the product is taken over all one-irreducible subgraphs $\chi_i$ (again including the graph $\chi$ as a whole) with the canonical dimension $n_i \geq 0$ and $a_i$ is  a parameter that stretches momenta flowing into the $i$-th subgraph inside this graph. The 
 main outcome of this approach is that integrals are finite for $\epsilon=0$.
This scheme enables us to calculate a contribution from each diagram to counterterms \cite{Adzhemyan2011}
\begin{equation}
Z_i = \frac{2}{n\epsilon}\left[ f_i - J \bar{\Gamma}_i^{(n)}\right],
\end{equation}
where $n$ is a number of loops. The second term on the RHS stands for a sum of diagrams of lower order perturbation theory. This allows us to  recursively calculate counterterms in the NM scheme at the normalized point and thus
gives us an opportunity to compare the results with ones obtained in the MS scheme. 
{\section{Calculation of anomalous dimensions} \label{sec-2}}
In this part, we illustrate the method described in the previous section by its application to a specific diagram.
Let us consider a two-loop contribution to $f_3$ from (\ref{eq-f}) determined by the two-loop
  three-point diagram of $\Gamma_4$
\begin{figure}[h]
\centering
\includegraphics[width = 5cm]{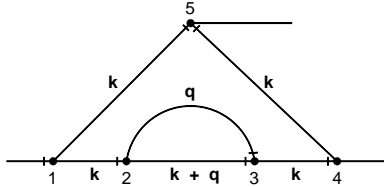}
\caption{The two-loop Feynman diagram of $\Gamma_4$ with symmetry factor equal to one.}
\end{figure}
with the dimension $n_\chi = 0$. The diagram has one relevant subgraph: the subgraph
given by vertices $\{2,3\}$ with the dimension $n_a = 2$.

The action of the differential operator $ - \tilde{\tau}  \partial_{\tilde{\tau}}$ on the line $G (k) = 1 /(k^2 +\tilde{\tau})$ produces an additional factor $1/(k^2 +\tilde{\tau})$ and it corresponds to the insertion of
a unit vertex into the propagator line. Graphically it will be denoted by the additional two-point 
 interaction vertex. The application of the operator to  the diagram results into 
a sum over all possible insertions of the vertex
\begin{equation}
 - \tilde{\tau}  \partial_{\tilde{\tau}} \chi = \raisebox{- 4 ex}{\epsfysize=2truecm \epsffile{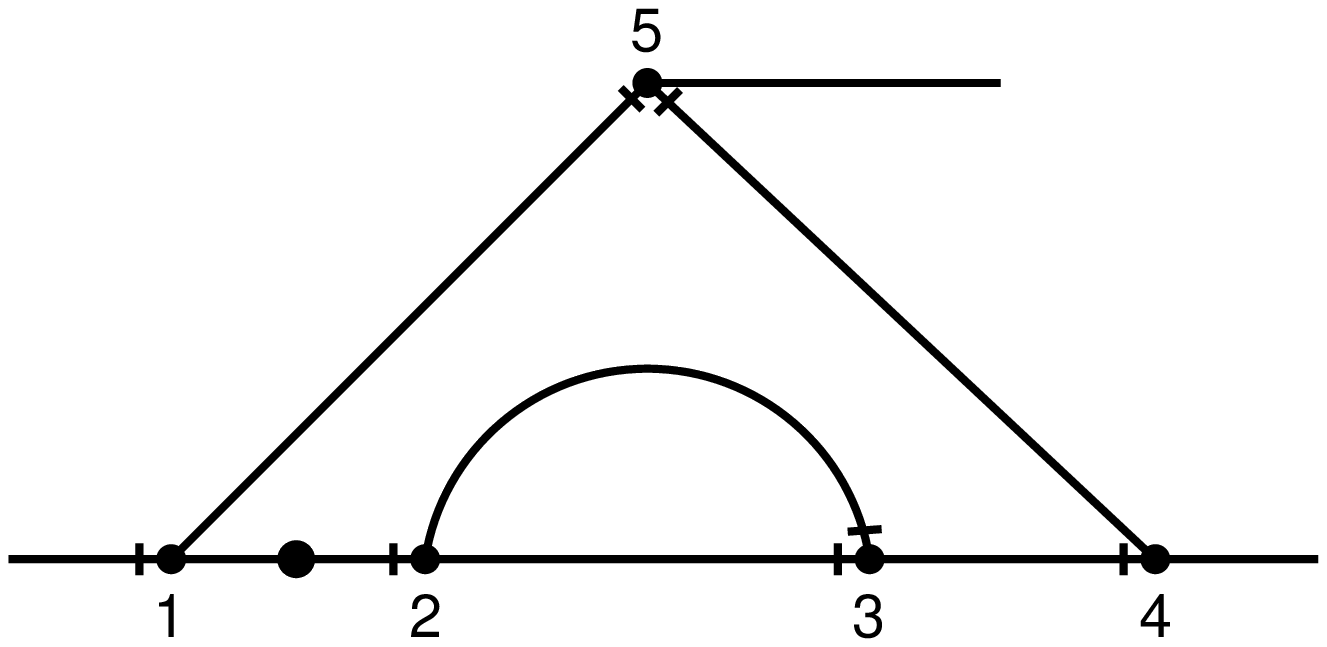}} +
\raisebox{- 4 ex}{\epsfysize=2truecm \epsffile{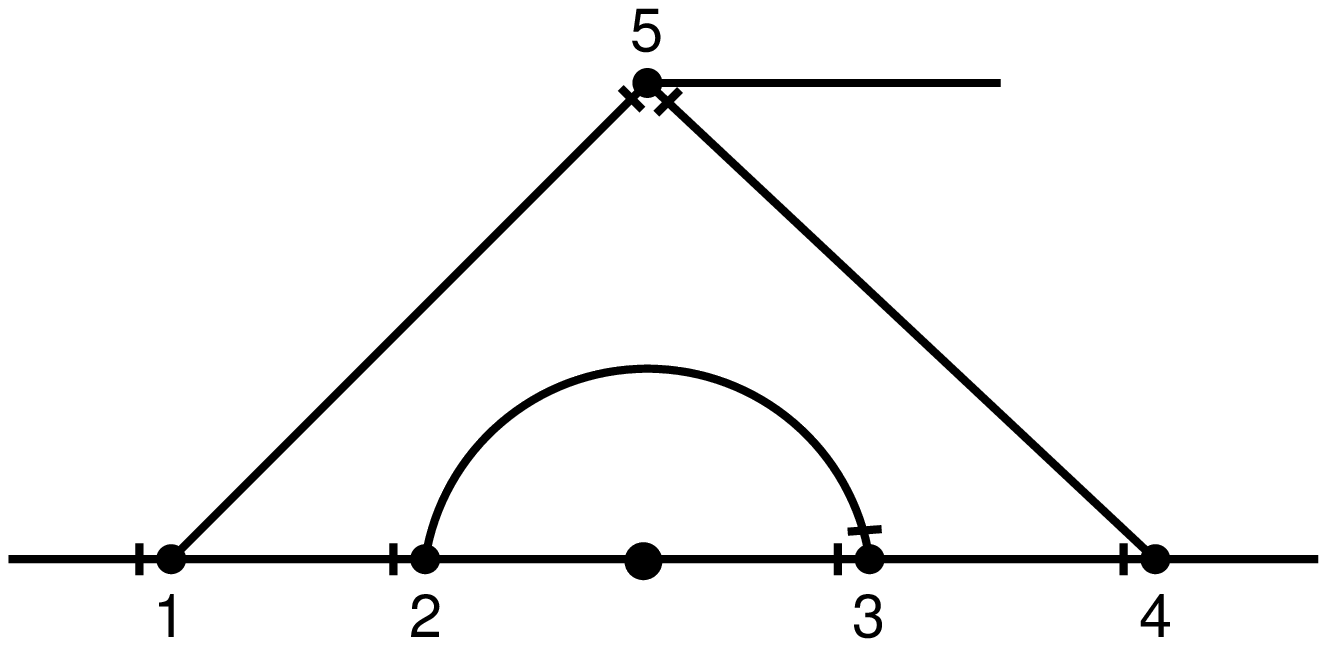}} + \dots  
 \label{eq:exp_chi}
\end{equation}

The next step consists of the inclusion of the operator $R$, but the analysis of each diagram has to
be made separately. For example for the first diagram on the RHS of Eq.(\ref{eq:exp_chi}), with the 
point outside the subgraph, nothing changes the dimension of the subgraph and $n_a = 2$ remains valid. On the other hand, 
 in the second graph, the subgraph becomes  logarithmic and the corresponding dimension
is changed to $n_a = 0$. In this way we obtain the expansion
\begin{equation}
R \left(- \tilde{\tau}  \partial_{\tilde{\tau}} \chi \right) \big|_{\tilde{\tau} = 1} = \frac{1}{2} \int\limits_{0}^{1} da (1-a)^2 \partial_a^3 
\raisebox{- 4 ex}{\epsfysize=1.8 truecm \epsffile{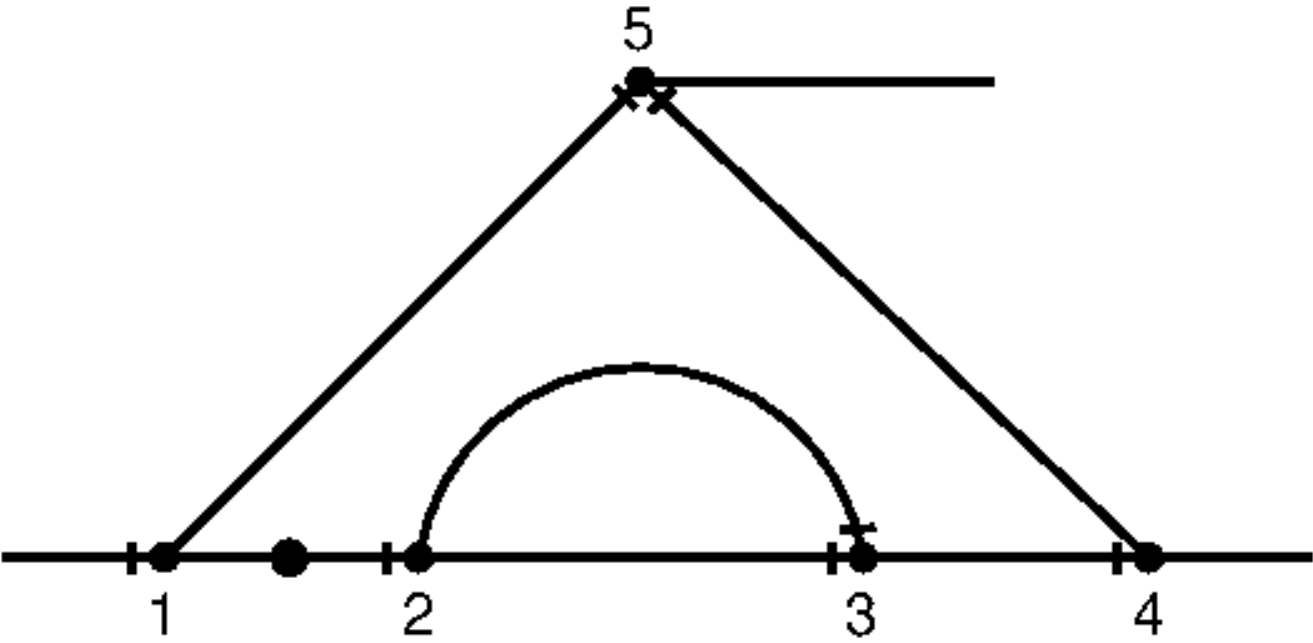}} + \int\limits_{0}^{1} da \partial_a
\raisebox{- 4 ex}{\epsfysize= 1.8 truecm \epsffile{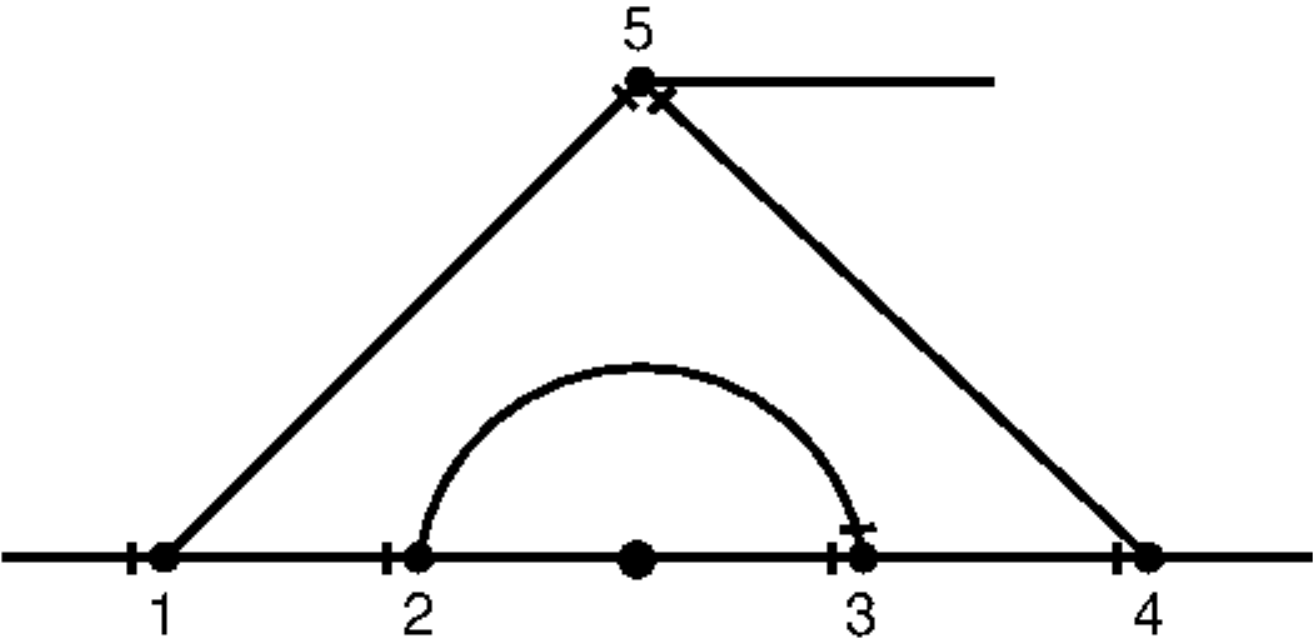}} + \dots. 
\end{equation}
Further it is necessary to multiply all external parameters for a given subgraph by the
 parameter $a$. For the propatagor line this means that $G (q + k) = 1 /[(a q+k)^2 + 1]$.

Combining (\ref{eq-gamma}) and (\ref{eq-f})  we can derive relations for anomalous dimensions $\gamma$ for fields and parameters of the model
\begin{align}
  &\gamma_{\psi} =\frac{f_1}{1+f_2}, 
  &\gamma_D& = \frac{2(f_2-f_1)}{1+f_2}, 
  &\gamma_{u}& = 2 \frac{2 f_4-f_1-2f_2}{1+f_2 },
  \label{eq:anomal_dim}
\end{align}
where $f_i$s are given up to the two-loop approximation by the following expressions:
\begin{equation}
  f_1 = -\frac{u}{16} +\frac{u\epsilon}{32} + 0.0152772 u^2, \quad 
  f_2 = -\frac{u}{32} +\frac{u\epsilon}{64} + 0.0062804 u^2, \quad 
  f_4 = -\frac{u}{4}  +\frac{u\epsilon}{8}  + 0.117185 u^2. 
  \label{eq:results_f}
\end{equation}
These results were obtained by a numerical calculation in which the actual form of integrals is determined by the
 $R$ operator using the Feynman representation.  Subsequently,  the momentum integrals are calculated
  by the Monte Carlo methods \cite{Cuba}. 
To the second order of perturbation theory, there are $2$ diagrams for 
the function $\Gamma_{\psi^{\dagger}\psi}$  and $11$ diagrams
for the function $\Gamma_{\psi^{\dagger} \psi^2}$. 

The scaling regimes are associated with the fixed points (FPs) of the RG transformation. The asymptotic large
scale behavior is governed by the infrared fixed points. Their coordinates can be found from
 the requirement that $\beta$-functions vanish. The directed bond percolation process has only one $\beta$-function
\begin{equation}
\beta_{u} = u ( -2 \epsilon - \gamma_u ) \approx u \left( - 2 \epsilon + \frac{3 u}{4} - \frac{3u\epsilon}{8} - \frac{3u^2\epsilon}{128} - 0.389626u^2 \right).
\end{equation}
There are two FPs given by the equation above:
 the trivial (Gaussian or free) FP ($u=0$) and the non-trivial one of the following form:
\begin{equation}
u^* = \frac{8}{3}\epsilon + 5.02756 \epsilon^2 + O(\epsilon^3).
\end{equation}
that corresponds to the critical percolation process.
After the determination of the FP coordinates, critical exponents can be analyzed.
 First, the critical exponent $\eta$ takes the following value
\begin{equation}
  \eta \equiv 2 \gamma_{\psi}\big|_{u=u^*} = - \frac{\epsilon}{3} -0.27228 \epsilon^2 + O(\epsilon^3).
  \label{eq:res_eta}
\end{equation}
The second is the so-called  dynamical exponent $z$ which is associated with the distinctive
behavior with respect to the time direction
\begin{equation}
  z \equiv 2-\gamma_D \big|_{u=u^*} = 2 - \frac{\epsilon}{6} -0.11682 \epsilon^2 + O(\epsilon^3).
  \label{eq:res_z}
\end{equation}
Needed momentum integrals were calculated with the numerical precision of $10^{-4}$.
%
%
For comparison with an analytic calculation we report the appropriately changed results (rescaling 
$\epsilon\rightarrow 2\epsilon$ is needed) from \cite{Janssen81,Janssen} 
%
%
\begin{eqnarray}
\eta &=& -\frac{\epsilon}{3}\left[1+ \left( \frac{25}{144} +\frac{161}{72} \textrm{ln} \frac{4}{3} \right) \epsilon + O(\epsilon^2) \right] 
\approx -\frac{\epsilon}{3} - 0.2723000633 \epsilon^2 + O(\epsilon^3), \\
z &=& 2-\frac{\epsilon}{6} \left[1+ \left( \frac{67}{144} +\frac{59}{72} \textrm{ln} \frac{4}{3} \right) \epsilon + O(\epsilon^2)\right] 
\approx 2 - \frac{\epsilon}{6} - 0.1168362090 \epsilon^2 + O(\epsilon^3).
\end{eqnarray}
We thus obtain excellent agreement with our results (\ref{eq:res_eta}) and (\ref{eq:res_z}). 

 Our two-loop results are in agreement with analytic calculations and our numerical method is suitable for the calculation
 of the  Feynman graphs to the three-loop order. To this end, it is necessary
 to take into account altogether $17$ graphs for the one-irreducible
  Green function $\Gamma_{\psi^{\dagger}\psi}$ and $150$ graphs for the function
  $\Gamma_{\psi^{\dagger} \psi^2}$.  It is also feasible to use this  method  
  for higher-loop computations. One just has to keep in mind that in order to achieve the required accuracy,
  computer time needed for calculation of each of the diagrams is much longer. The work in this direction
  is still in progress.
\begin{acknowledgement}
This work was supported by VEGA grant 1/0222/13.  L.~Ts.~A. and M.~V.~K. acknowledge the Saint Petersburg State University for
the research grant 11.38.185.2014. We would also like to thank Dr. M. Vala  the coordinator of the project ``Slovak infrastructure 
for high performance computing (SIVVP), ITMS 26230120002''.
\end{acknowledgement}
%
%
%

\end{document}